# Kinetomagnetism of chirality and its applications

Sang-Wook Cheong[1,a] and Fei-Ting Huang[1]

[1]*Rutgers Center for Emergent Materials and Department of Physics and Astronomy, Rutgers University, Piscataway, NJ 08854, USA*

[a] Author to whom correspondence should be addressed. Electronic mail: sangc@rutgers.physics.edu

**Chiral functionalities exhibited by systems lacking any mirror symmetry encompass natural optical activity, magnetochiral effect, diagonal current-induced magnetization, chirality-selective spin-polarized current of charged electrons or neutral neutrons, self-inductance, and chiral phonons. These phenomena are unified under the hypothesis of 'kinetomagnetism of chirality', which posits that any moving (charged or neutral) object in chiral systems induces magnetization in its direction of motion, consequently imparting chirality to the object due to this induced magnetization. We also found conjugate relationships among the 'kinetomagnetism of chirality', 'linear magnetoelectricity', and 'electric field induced directional nonreciprocity', highlighting their interconnections with magnetic, electric, and toroidal orders. The concept of the kinetomagnetism of chirality will be an essential basis for the theoretical understanding of known chiral phenomena such as natural optical activity or chiral phonons, and also the discovery of unexplored chiral functionalities.**

## I. Introduction

Magnetic fields generated in a coil or solenoid with a twisted structure epitomize the intricate relationship between electricity and magnetism as outlined in the Maxwell's equations. This interplay is fundamental to contemporary technologies, including microelectronics, electric car motors and power generation. Intriguingly, certain conductive crystalline solids exhibit nanoscale twisting. This raises the question: can these twisted crystals produce magnetic fields and demonstrate inductance when subjected to macroscopically uniform electric currents, whether constant or varying? The answer is affirmative, revealing a remarkable attribute inherent to these twisted crystals—chirality[1-3].

Many objects in nature, such as coils, screws, spiral stairs, and DNA, exhibit twisted structures, but twisting isn't a prerequisite for chirality. Chirality describes a property where an object and its mirror image cannot be superimposed, regardless of spatial rotations and translations[1-4]. This means that the object lacks mirror symmetry in any orientation or position. An electric coil or solenoid does have chirality: the mirror image of a coil twisted along one direction becomes a coil twisted along the other direction, and these two coils cannot be superimposable, no matter how they are rotated or positioned. Chiral molecules, on the other hand, often lack discernible twisting. Thus, the question expands further: can chiral crystals, even without apparent twisting, generate magnetic fields and exhibit inductance when subjected to macroscopically uniform electric currents?

Another intriguing question also arises: what occurs with induced magnetic fields or inductance when a magnetic state, like a helical spin state, disrupts all mirror symmetries[5, 6], even in the presence of an achiral crystallographic structure? Moreover, what unfolds in magnetic states when macroscopically uniform electric currents are introduced and both time reversal symmetry and all mirror symmetries are broken? The goal of this perspective is to address fundamental questions through a focus on symmetry, aiming for a succinct explanation of various chiral-related phenomena, such as natural optical activity, magnetochiral effects, self-inductance, and chiral phonons. Our hypothesis, 'kinetomagnetism of chirality', wherein the prefix 'kineto-' referring to motion or movement, highlights the connection between the motion of chiral objects and the generation of induced magnetism. Through the lens of 'kinetomagnetism of chirality', we explore diverse chiral functionalities, conjugate properties, and potential candidates using the framework of magnetic point groups[7].

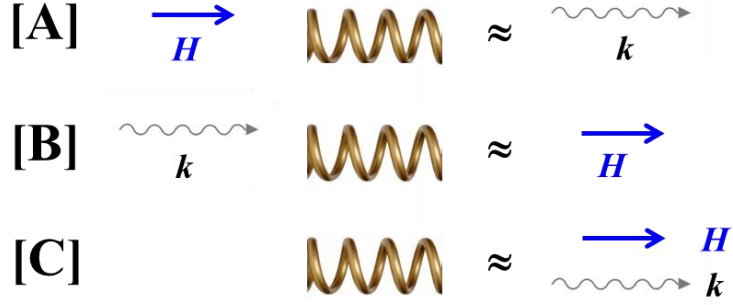

**FIG. 1.** Three permutable SOS relationship among chirality (represented by golden-color springs), constant velocity (linear momentum, $k$) and magnetic field ($H$).

## II. Kinetomagnetism of chirality

Kinetoelectricity, as described by Ascher (1974)[8] and Schmid (2008)[9], refers to the generation of electricity within a moving crystal, directly proportional to its velocity, and in the absence of any externally applied electromagnetic field. Similarly, kinetomagnetism describes a distinct concept or phenomenon involving the conversion or interaction between moving entities and magnetism, separate from the reported instances of kinetic magnetism discussed in various works[10, 11]. Here, a hypothesis termed 'kinetomagnetism of chirality' describes that any (charged or neutral) object moving along a given direction in chiral systems induces magnetization ($M$) in that same direction, which results in the object itself acquiring chirality due to the induced magnetization. This hypothesis is derived from three permutable Symmetry Operational Similarity (SOS)[12-14] relationships pertinent to chirality shown in Fig. 1 [A]-[C], where chirality, constant velocity (linear momentum), and magnetic field are represented by a golden-color spring, $k$ and $H$, respectively. The symbol '≈' distinguishes between the left-hand side, representing 'a sample subjected to various environments such as applied electric fields, magnetic fields, or electric currents,' and the right-hand side, representing 'measurable aspects of a phenomenon.' The environments on the left-hand side exhibit equal or more broken symmetries compared to the measurable aspects of the phenomenon on the right-hand side, enabling the occurrence of the phenomenon. For example, the spring may have only 3-fold rotation ($C_3$) symmetry along the horizontal direction, but the right hand-sides do have ∞ rotational symmetry.

**A. Magnetochiral effect**

First, Fig. 1[A] tells us that a chiral system in the presence of ***H*** has SOS with ***k***. If so, any symmetry operation inducing the 180° flipping of ***k*** (i.e., +***k*** becomes -***k***) will also change the left-hand side – for example, change +***H*** to -***H*** or switch left chirality to right chirality. What it means is that the behavior of +***k*** in the left-hand side can be different from that of -***k*** in the same left-hand side. In other words, there can be Nonreciprocal Directional Dichroism (NDD)[5, 15-17], which is called a magnetochiral effect[18-20] in the case of chiral systems. This magnetochiral effect has been observed for electronic transport as well as propagation of acoustic phonons[18], spin waves[21], and light[16]. Figure 2(a-b) exemplifies the nonreciprocal directional dichroism in chiral multiferroic $Ni_3TeO_6$ single crystal. $Ni_3TeO_6$ crystallizes in a polar and chiral *R*3 structure with collinear antiferromagnetic (AFM) ordering along the *c* axis below $T_N$=52 K (MPG 31′). Reversing the light direction of *k* (Fig. 2(b) red and blue curves) while keeping magnetic field H, distinct absorption difference curves ($\Delta\alpha_{NDD}$, green curve in Fig. 2(b)) are observed[16, 17].

**B. Electromagnetic induction**

Figure 1[B] suggests that a constant motion with ***k*** in a chiral system has SOS with ***H***. This SOS relationship indicates that macroscopically uniform DC electric current in a chiral system can induce ***H***. This was theoretically first predicted in 2015 [22] and then experimental observation was reported in 2017 [23]. Figures 2(c-d) demonstrates the observation of current-induced electronic magnetization in bulk tellurium using nuclear magnetic resonance (NMR) measurements. Te crystallizes in a chiral structure with space group (SG) $P3_121$. The distinct responses of current-induced electronic magnetization are clearly observed when applying pulsed currents of 0 and ± ***k*** as depicted in black, red, and blue signals. Note that a much earlier report of linear change of optical rotation of transmitted THz light in chiral Te [24] in the presence of DC electric current appears in accordance with this SOS relationship. Under this [B]

SOS relationship, time-varying current naturally induces time-varying **H**, which could result in self-inductance[25].

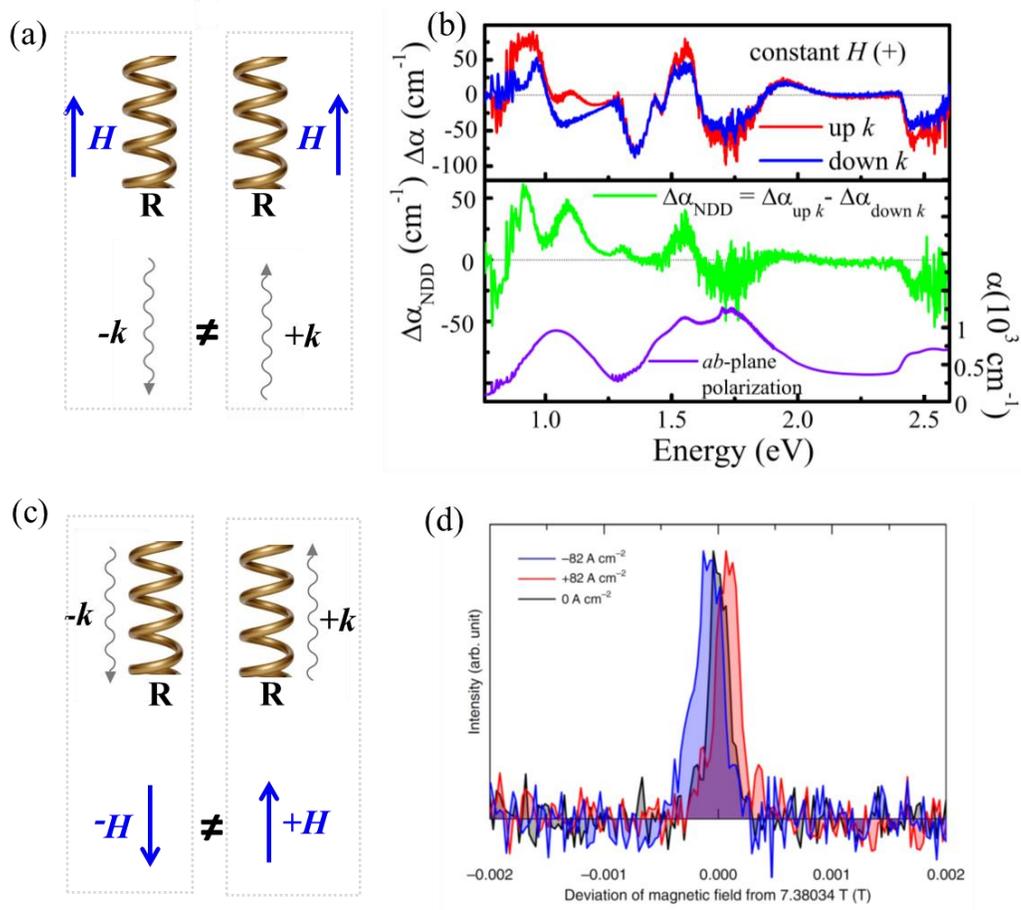

**FIG. 2** (a) Schematics of **H** and **k** relative to a right-handed domain in the magnetochiral measurement. (b) Nonreciprocal directional dichroism between +**k** and -**k**, measured in chiral multiferroic $Ni_3TeO_6$ with the H // k // [001] configuration at 4 K. Blue, red, and green curves represent absorption difference Δα along +**k**, -**k**, and the difference between the absorption along these two directions, respectively. Reproduced with permission from Yokosuk *et al.*, npj Quantum Mater. 5, 20 (2020). Copyright 2020 Springer[17]. (c) Schematics of **H** and **k** relative to a right-handed domain in the current-induced magnetization measurement. (d) Current induced magnetization in a chiral Te single crystal. The black, red, and blue lines correspond to data obtained under pulsed electric current densities of 0, +82, and −82 $Acm^{-2}$, respectively. The shift in line position in the presence of an applied current clearly indicates the emergence of electronic magnetization. Reproduced with permission from Furukawa *et al.*, Nat. Commun. 8, 954 (2017). Copyright 2017 Springer[23].

## C. Chiral phonon

Figure 1[C] indicates that a combination of *k* and *H* can act like chirality. Numerous phenomena can be understood in terms of this simple statement. For instance, in 1956, the asymmetry observed in electron propagation during the β-decay of $Co^{60}$ in the presence of *H* can be understood as the breaking of chiral (mirror) symmetry. This asymmetry arises from the combined effect that mimics chirality, consequently resulting in the breaking parity symmetry[26]. Some argue that the enantiomeric imbalance in biological systems on earth originates from the preferred chirality due to the simultaneous presence of sun light propagation and earth magnetic fields[27].

Moreover, the Schwinger scattering[28-30] also stems from the Figure 1[C] SOS relationship. Polarized neutron scattering has been typically used to unveil the exact magnetic ordering configurations. Recently, polarized neutron scattering has been demonstrated to reveal the absolute chirality of helical spin state or crystallographic chirality of non-magnetic chiral systems, and this is called Schwinger scattering. In other words, spin-polarized neutron can have chirality and the coupling of this neutron chirality and specimen chirality can result in the Schwinger scattering. Similarly, spin-polarized tunneling electric current can couple with chirality of a conducting specimen with no magnetic ordering, which is referred to as 'chiral tunneling'[31, 32]. In chiral tunneling within a chiral conducting system, the tunneling electric current induces *M* along the tunneling current direction. The sign of this induced magnetization depends on the chirality of the system. This induced *M* couples with spin polarization of the tunneling current, facilitating chirality-selective tunneling within the chiral system.

Natural optical activity[33-35] of chiral molecules, i.e. the rotation of the polarization direction of linearly polarized light when the light travels through a chiral material, was discover in late 19 century, and Lod Kelvin[3] coined the term of 'chirality' to describe this phenomenon. This natural optical activity as well as all chiral phenomena discussed in conjunction with Figure 1[A]-[C] SOS relationships can be ubiquitously understood under a hypothesis of the 'kinetomagnetism of chirality', i.e. any (charged or neutral) object moving along a given direction in chiral systems induces magnetization (*M*) in that same direction, which results in the object itself acquiring chirality due to the induced magnetization. In the case of natural optical activity, moving light in a chiral system induces *M* through the "kinetomagnetism of chirality" and this induced *M* couples with light polarization, so results in natural optical activity. Since the induced *M* changes its sign when light *k* flips, so this natural optical activity is reciprocal.

Any quasi-particle motion in a chiral system induces *M*, and this induced *M* naturally couples with applied *H*, and this coupling induces, for example, the magnetochiral effect. Evidently, the induced *M* changes its sign for the opposite chirality, so the magnetochiral effect depends on chirality. The current-induced *H* in chiral systems associated with the Fig. 1[B] SOS relationship and also the numerous phenomena associated with the Fig. 1[C] SOS relationship, including chiral symmetry breaking in the β-decay of $Co^{60}$, the chirality control of helical spin states, a scenario for the enantiomeric imbalance in biological systems, Schwinger neutron scattering, and chiral electric tunneling, and chiral phonons[36-38], are clearly in accordance with the 'kinetomagnetism of chirality'.

**D. Kinetomagnetism of chirality: Applications**

Through kinetomagnetism of chirality, anything moving with non-zero *k* in chiral systems becomes chiral through induced *M*, so any quasi-particles with non-zero *k* in chiral systems such as optical or acoustic phonons, magnons, electromagnons, electrons, holes, light, *etc*. naturally become chiral through induced *M*. Thus, phonons with non-zero *k* in chiral system can be associated with a magnetic character and naturally become chiral phonon. A probe with a chiral nature, such as circularly polarized light known for its helical wavefronts and the rotation of the electric field vector either clockwise or counterclockwise, can effectively couple with chiral phonons. This coupling has been observed in materials like cinnabar α-$HgS$[36] and quartz[37], in which both crystalize in chiral magnetic point group (MPG) 32. Chiral magnetic point group requires the breaking of space inversion (denoted as **I**) and the combined symmetry breaking of space and time inversion (symbolized as **I⊗T**) when the crystal can freely rotate (FR)[2, 5]. The 𝒞 chiral (diagonal kinetomagnetic point group consist of 32 MPGs: 1, 11', 2, 21', 2', 222, 2221', 2'2'2, 3, 31', 32, 321', 32', 4, 41', 4', 422, 4221', 4'2'2, 42'2', 6, 61', 6', 622, 6221', 6'22', 62'2', 23, 231', 432, 4321', 4'32'.[5] In materials with those chiral MPGs or magnetic states showing chirality[5], such as incommensurate helical orders, Bloch-type ferromagnetic walls, and Bloch-type skyrmions, (acoustic) phonons or thermal currents can acquire intrinsic chirality. A number of exemplary spin-ordered states with $\mathcal{C}$-type MPGs are presented in Fig. 3(a-d). A helical spin state (Fig. 3(a)), a magnetic toroidal moment combined with out-of-plane dipole (*i.e.*, spin canting in Fig. 3(b)), a magnetic quadrupole combined with magnetic octupole (*i.e.* alternating spin canting in Fig. 3(c)), and a Bloch-type skyrmion (Fig. 3(d)) demonstrate the $\mathcal{C}$ characteristics. Figs. 3(b), 3(c), and 3(d) correspond to 42'2', 4'2'2, and 42'2', respectively.

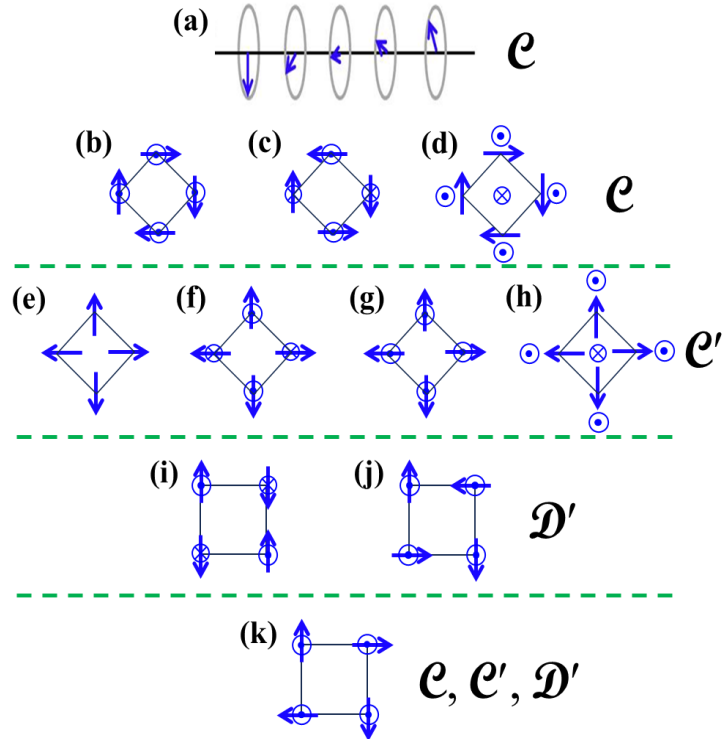

**FIG. 3.** Various magnetic states belonging to the $\mathcal{C}$, $\mathcal{C}'$, or $\mathcal{D}'$ point groups. (a) depicts a helical spin state. (b-d) correspond to MPGs 42'2', 4'2'2, and 42'2', respectively. (a-d) are $\mathcal{C}$-type. (e-h) correspond to MPGs 4/m'm'm', $\bar{4}$'m'2, 4m'm', and 4m'm', respectively, and are $\mathcal{C}'$-type. (i-j) correspond to MPGs 2/m and $\bar{4}$, respectively, and are $\mathcal{D}'$-type. (k) corresponds to MPG 4, which can be any of $\mathcal{C}$-type, $\mathcal{C}'$-type, and $\mathcal{D}'$-type. Note that tetramerized square lattices are considered.

In these chiral materials, phonon angular momentum can be generated by the temperature gradient, known as phonon thermal Edelstein effect[39], which holds potential for generating spin currents in response to thermal gradients[40]. Additionally, even in materials with nonchiral MPGs (including $\bar{4}$, $\bar{4}1'$, $\bar{4}'$ $\bar{4}2m$, $\bar{4}2m1'$, $\bar{4}2'm'$, $\bar{4}'2'm$, and $\bar{4}'2m'$)[41], diagonal current-induced magnetization can occur. We predict that this phenomenon could potentially reveal chiral phonons through the concept of kinetomagnetism of chirality. Among the nonchiral MPGs exhibiting diagonal current-induced magnetization, only $\bar{4}1'$ and $\bar{4}2m1'$ are nonmagnetic and chiral phonons may be expected in compounds in nonmagnetic states. Semiconductor chalcopyrite family, such as $CuInTe_2$ [42] and $CuFeS_2$ [43], are of interest. $CuInTe_2$ is a non-magnetic p-type semiconductor that crystallizes in the targeted structure with SP $I\bar{4}2d$ (MPG $\bar{4}2m1'$). $CuFeS_2$ with the same $I\bar{4}2d$ structure has additional magnetic ordering ($T_N$=823K) corresponding to MPG $\bar{4}2m$. This suggests opportunity for demonstrating diagonal current-induced magnetization and observing chiral phonons above and below the magnetic ordering phase.

Furthermore, through the idea of kinetomagnetism of chirality, the simultaneous presence of mutually parallel electric current (**k**) and magnetic field (**H**) can induce chiral behavior, which can be utilized to control magnetic domains exhibiting chiral characteristics. For example, MnP forms in a centrosymmetric lattice with SG *Pnma* and show multiple magnetic transitions. The achiral to chiral transition comes in between 55-60 K[44, 45] and thus, there exist left-chiral and right-chiral helical magnetic domains. The spin helicity degeneracy is lifted in the presence of magnetic field **H** and electric current **j**, depends on whether they are parallel or antiparallel to each other. For example, applied ($+H_p$ $+j_p$) or ($-H_p$ $-j_p$) can favor the left-chiral helical magnetic domain, and ($+H_p$ $-j_p$) or ($+j_p$ $-H_p$) stabilizes the right-chiral helical magnetic domain as depicted in schematics of Fig. 4(a). Figures 4(b) demonstrate distinct electrical resistivity after various poling histories with combinations of magnetic fields $H_p$ and dc electric currents $j_p$[45] in MnP. Similar manipulations of helical magnetic domains should work in other materials such as kagome metal $YMn_6Sn_6$ with $T_{hel}$ = 326K[46], $GdRu_2Si_2$[47], $Mn_3Si$ with $T_{hel}$ = 25.8K[48], and $Gd_3Ru_4Al_{12}$ with $T_{hel}$ = 17.2K[49].

On the other hand, in insulating helical materials such as RbFe(MoO$_4$)$_2$ with a centrosymmetric lattice (SG $P\bar{3}$) and a helical order at $T_{hel}$ = 3.8K (MPG 31') [50, 51], thermal current offers an alternative source to manipulate magnetic domains.

Moreover, in materials exhibiting chiral crystallographic lattices, both thermal current and electric current can induce magnetization even in their paramagnetic states. This induction has potential to trigger the switching of the net magnetic moment direction, a spin-flop transition, or change ferromagnetic/antiferromagnetic domains below magnetic transition temperatures. Such phenomena can be tested, for example, for various magnetic states switching in chiral (SG $P6_322$) Fe$_{1/3\pm\delta}$NbS$_2$ at $T_N$ = 45K (MPG 2221')[52, 53].

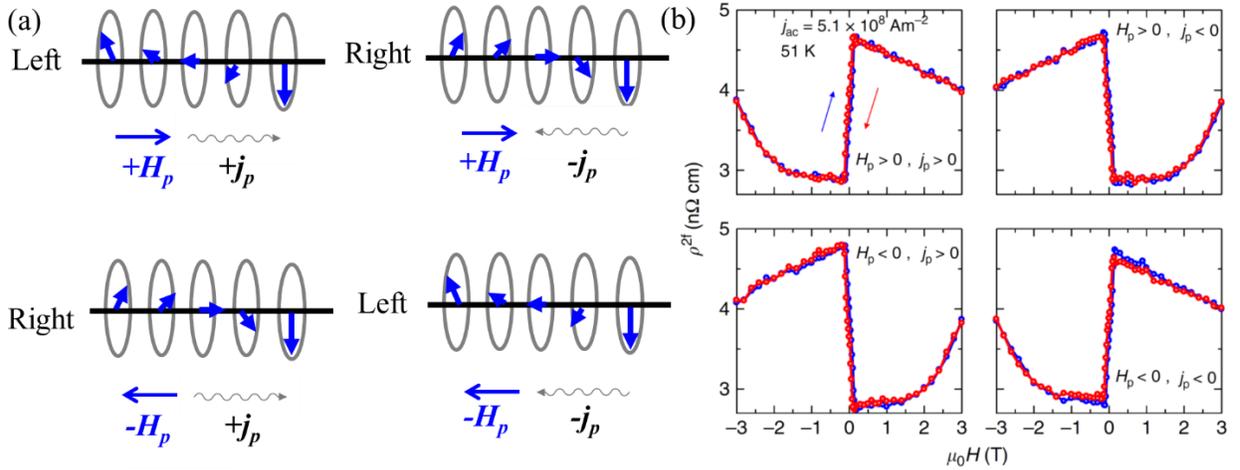

**FIG. 4** (a) The preferred helical domains are contingent upon the alignment of $H_p$ and $j_p$, determining their energy favorability. For example, a left-handed chiral domain is favored when both $H_p$ and $j_p$ are parallel (+$H_p$ +$j_p$, or -$H_p$ -$j_p$), while right-handed chiral domains are favored in antiparallel configurations (+$H_p$ -$j_p$ or -$H_p$ +$j_p$). (b) The 2$^{nd}$ harmonic contribution of electrical resistivity $\rho^{2f}$ and the proposed favored helical domains at 51 K in MnP crystals. As a function of the magnetic field along the propagation vector (crystallographic *a*-axis), the measurements were done after the different poling procedures as indicated. Reproduced with permission from Jiang et al., Nat. Commun. 11, 1601 (2020). Copyright 2020 Springer[45].

To explore local current-induced dynamics in magnetization/chirality reversal of chiral ferromagnetic domain walls[54] or chiral ferromagnet remain to be explored. Chiral ferromagnetic point group such as 1, 2, 2′, 3, 4, 6, 2′2′2, 32′, 42′2′ and 62′2′ could be potential candidates for this application such as LaMn$_2$Ge$_2$ showing large topological hall effect, which undergoes a transition from centrosymmetric lattice (SG $I4/mmm$) to a chiral ferromagnetic state with MPG 42′2′ at 322 K[55]. Pyrochole Yb$_2$Ge$_2$O$_7$ also crystallize in 42′2′ magnetic states[56]. Those chiral ferromagnetic point group should exhibit non-reciprocal optical effect[17] as explained in Section II A.

This hypothesis of "kinetomagnetism of chirality" is straightforward in the view of symmetry, and appears extremely general and powerful; however, it requires future rigorous theoretical understanding in terms of microscopic mechanism. The understanding this microscopic mechanism will be essential to optimized materials with enhanced chiral functionalities. Emphasize that the chiral functionalities can find broad applications in, for example, chiral spintronics[57] and chiral quantum optics[58].

**III. Kinetomagnetism of chirality: Conjugate properties**

Next, we explore the discussion on conjugate properties of kinetomagnetism of chirality as shown in Fig. 5. Symmetry-driven conjugate properties have been observed in magnetic (***M*** or ***H***), electric (***P*** or ***E***), and toroidal (***k***) modes, along with their relevant properties [9, 41, 59-61]. Magnetic, polar, and toroidal MPGs, which can interconvert, exemplified by transformation such as ***M***↔***P***, ***P***↔***k***, and ***k***↔***M*** through the **I**↔**T**, **T**↔**I⊗T**, and **I⊗T**↔**I** symmetry operation as depicted as green, blue, and red lines, respectively in Fig. 5. The top corner of Fig. 5 illustrates kinetomagnetism of chirality, which describes the three-fold nature involving 𝒞 (chirality), ***k*** and ***H***. This phenomenon breaks space inversion symmetry (**I**) and the combined symmetry of both space and time inversion (**I⊗T**), symbolized as {**I**, **I⊗T**}, where {} signifies broken symmetries.

On the other hand, at the right vertices of the triangle in Fig. 5, linear magnetoelectric effect is depicted. Linear magnetoelectric effect is highly appealing for potential devices since an external electric field induces a magnetic magnetization or vice versa. $Cr_2O_3$ stands out as the most known linear magnetoelectric material. $\mathcal{C}'$ (chirality prime) group has been defined as the **T** symmetry in addition to all mirror symmetries are broken, i.e. broken {**I**, **T**}, with free spatial rotations[5], exhibiting diagonal linear magnetoelectricity. Spin-ordered states with $\mathcal{C}'$-type MPGs are presented in Fig. 3(e), 3(f), 3(g), and 3(h), corresponding to MPGs $4/m'm'm'$, $\bar{4}'m'2$, $4m'm'$, and $4m'm'$, respectively. Similarly, a combination of **k** and **H** can act like chirality while the coupled electric and magnetic fields can act as $\mathcal{C}'$.

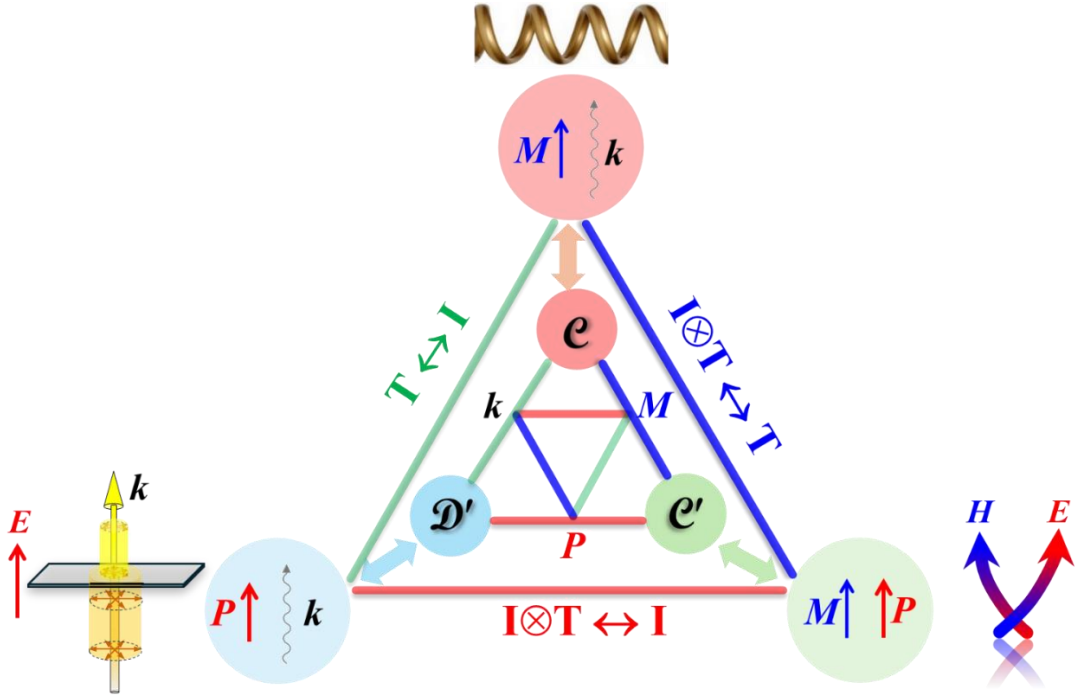

**FIG. 5** The symmetry-driven conjugate trinity of the "kineomagnetism of chirality" (red circles), linear magnetoelectricity & $\mathcal{C}'$ (green circles), and electric field induced directional nonreciprocal effect & $\mathcal{D}'$ (blue circles) are illustrated. The essence of conjugation highlights the three-fold nature of these colored circles, for example, $\mathcal{C}$, **k** and **H** in red circles. The routes of symmetry interconversion between these phenomena are depicted by blue, red, and green lines, respectively, aligning with the three known conjugate ferroic orders–magnetic (**M**), electric (**P**), and toroidal (**k**), located at the center.

Conversely, kinetomagnetism of chirality vanish when inversion **I** is conserved while both time inversion **T** and their product **I⊗T** are broken as {**T**, **I⊗T**}. This category is associated with the electric field-induced directional nonreciprocity at the left vertices of the triangle in Fig. 5. Unlike the classic Faraday effect where a magnetic field induces nonreciprocal effects, an electric field (represented by red arrows) can induce a similar electro-optic effect in materials with broken {**T**, **I⊗T**}, i.e. $\mathcal{D}'$ (director prime) materials[62]. α-$Mn_2O_3$ [63] and $Ca_2RuO_4$ [64] are materials belongs to the $\mathcal{D}'$ [62] group, which has been introduced as electric-field tunable toroidal moments[62]. Spin-ordered states with $\mathcal{D}'$-type MPGs are presented in Figs. 3(i)-(j), corresponding to $2/m$ and $\bar{4}$, respectively. Similarly, the conjugate relationships indicate that applying an electric field on a $\mathcal{D}'$ group materials along a direction can result in nonreciprocal effect along the same direction, while applying (electric) current on a $\mathcal{D}'$ group materials along a direction can induce electric polarization along the same direction. Note that there are eleven MPGs (1, 2, 222, 3, 32, 4, 422, 6, 622, 23, 432) involving $\mathcal{C}$, $\mathcal{C}'$, and $\mathcal{D}'$ and also all properties discussed. The spin state depicted in Fig. 3(k) corresponds to the MPG 4, which can be any of $\mathcal{C}$, $\mathcal{C}'$, and $\mathcal{D}'$.

**IV. Conclusion**

Chirality, characterized by all broken mirror symmetries, give rise to numerous exotic physical phenomena such as natural optical activity, the magnetochiral effect, diagonal current-induced magnetization, chirality-selective spin-polarized currents for charged electrons or neutral neutrons[65], self-inductance, and chiral phonons. These phenomena find succinct explanation through the hypothetical concept of 'kinetomagnetism of chirality', rooted in symmetry consideration. Under the concept of the 'kinetomagnetism of chirality', any motion of quasi-particles (electrons, phonons, magnons, electromagnons, *etc*.) or light in chiral systems become chiral through induced magnetization along its direction of motion. Applications of the concept of the 'kinetomagnetism of chirality' include the theoretical understanding of known chiral phenomena such as natural optical activity or chiral phonons. Additionally, it unveils chiral functionalities such as controlling ferromagnetic/antiferromagnetic domains, dynamics induced by local currents in magnetization/chirality changes and observing chiral phonons in systems conventionally regarded as nonchiral.

Furthermore, we have explored the symmetry-driven conjugate trinity among 'kineomagnetism of chirality', 'linear magnetoelectricity & $\mathcal{C}'$', and 'electric field induced directional nonreciprocity & $\mathcal{D}'$'. The essence of conjugation highlights the inherent three-fold nature in each phenomenon and interconnectedness of three primary ferroic orders: magnetic, electric, and toroidal. These symmetry-guided perspectives of chirality not only deepen our understanding of seemingly distinct properties but also open possibilities for discovering fresh or enhanced chiral functionalities, both theoretically and experimentally.

## ACKNOWLEDGMENTS

This work was supported by the W. M. Keck foundation grant to the Keck Center for Quantum Magnetism at Rutgers University.

## AUTHOR DECLARATIONS

### Conflict of Interest

The authors have no conflicts to disclose.

### Author Contributions

S.W.C. conceived and supervised the project. F.-T.H. conducted magnetic point group analysis.

## DATA AVAILABILITY

The data that support the findings of this study are available from the corresponding authors upon reasonable request.